\documentclass[psfig]{mn2e}
\usepackage{graphicx}
\newcommand{\e}{\begin{equation}}
\newcommand{\dd}{\mathrm d}
\newcommand{\st}{\end{equation}}
\newcommand{\m}{\begin{displaymath}}
\newcommand{\n}{\end{displaymath}}

\title{Electromagnetic Fields in Jets}

\author[B. D. Sherwin and D. Lynden-Bell]{B. D. Sherwin$^{1,2}$ and D. Lynden-Bell$^{2,3}$ \\\\
$^{1}$ Trinity Hall, Trinity Lane, Cambridge CB2 1TJ\\
$^{2}$ Institute of Astronomy, The Observatories, Cambridge CB3 0HA\\
$^{3}$ Clare College, Trinity Lane, Cambridge CB2 1TL}
\date{Accepted 28 March 2007}

\pagerange{1--8}
\pubyear{2007}
\begin{document}
\label{firstpage}

\maketitle

\begin{abstract}
The magnetic fields and energy flows in an astronomical jet described by our earlier model are calculated in detail. Though the field distribution varies with the external pressure function $p(z)$, it depends only weakly on the other boundary conditions. Individual fieldlines were plotted; the lines become nearly vertical at the bottom and are twisted at the top. An animation of a fieldline's motion was made, which shows the line being wound up by the accretion disc's differential rotation and rising as a result of this. The distribution of Poynting flux within the jet indicates that much of the energy flows up the jet from the inside of the accretion disc but a substantial fraction flows back down to the outside.
\end{abstract}

\section{Introduction}
Collimated radiating jets are a feature of many astronomical objects. They have been observed in radio galaxies (see Hargrave and Ryle (1974), Carilli and Barthel (1996), Gizani and Leahy (2003), Begelman, Blandford and Rees (1984)), quasars and young stellar objects. They may also be present in gamma ray bursts (see Uzdensky and MacFadyen (2006)), dying stars and microquasars (see Mirabel and Rodriguez (1999)). In all these objects, the jet emerges along the axis of an accretion disc. As the
form of the highly relativistic jets in active galactic nuclei and the much slower ones around stars is very similar, it is likely that relativistic effects are not essential for the formation of jets. The presence of strong magnetic fields in radio galaxies was deduced from their synchrotron radiation by Baade (1956), but the similarity of the observed structures suggests that magnetism plays a large role in the formation of all jets. In particular, recent observations of maser emission confirm that magnetic fields are important in collimating stellar jets, i.e. Vlemmings et al. (2006).

Following earlier numerical work on the Grad-Shafranov equation by Barnes and Sturrock (1972) and Sturrock and Barnes (1972), Lovelace (1976) obtained jet-like magnetic structures by considering the winding up of a pre-existing uniform field. However, collimation by a pre-existing field side-steps the problem of why the field is collimated to start with. In Lynden-Bell (1996) (paper II), the growing magnetic towers picture was developed in which a magnetic field twisted by an accretion disc was confined by an ambient coronal pressure. This provided a crude picture as to why the field was collimated. In later work (Lynden-Bell (2006) - paper IV) a non-relativistic model was developed which involved a force-free magnetic field being twisted up by a swirling accretion disc to give a jet-like collimated magnetic tower, now in a stratified atmosphere. Expressions for the magnetic fields were given; however, the fields were only illustrated in the idealised ``dunce's cap'' configuration for which the solutions were explicit and exact. In this paper the more general implicit expressions will be numerically evaluated to depict in detail the fields and energy flows within the jet predicted by the model under various boundary conditions.

Kato et al. (2004) have computed the starting of such jets in a magnetic accretion disc. Li et al. (2006) extended these ideas to long jets that are not force-free initially but become so as they evolve. Nakamura, Li and Li (2006) performed numerical calculations of jets within an isothermally stratified atmosphere up to heights that are twice the diameter of the disc; our analytical solutions extend to far greater heights. Discussions of jet stability are given by Nakamura and Meier (2004) and Nakamura, Li and Li (2007). Lovelace et al. (2002) computed solutions in a toroidal box and found a strongly collimated Poynting flux as predicted by the growing towers picture. However, their calculations were limited by their box. Romanova et al. (1998) and Ustyugova et al. (1999) computed the evolution of magnetic fields above discs, with the fields not force-free. Though Ustyugova et al. considered fields initially of split monopole type and Romanova et al. treated a more complicated field structure, both obtained magnetically dominated collimated outflows with the matter flow everywhere away from the disc. As discussed later, we find that there is also a smaller Poynting flux back down onto the outer feet of the fieldlines, which is due to the fieldlines' tension pulling on the disc. There are many simulations of highly relativistic black hole environments, such as Gammie et al. (2003), McKinney (2005) and McKinney and Narayan (2007a/b).\\ \\ \\

\section{Theoretical Basis}
A brief summary of the theory in paper IV will be provided, as the investigation of the fields which follows is based on the model developed in this paper. Cylindrical polar coordinates are used. The magnetohydrodynamic equations are assumed in mixed Gaussian units.

\subsection{Introducing the Problem to be Solved}
The field is located within a magnetically dominated cavity in the external medium. The displacement current is neglected, so that $\nabla \times \mathbf B= 4 \pi \mathbf j$. Though the inside of the cavity has negligible gas pressure, it is assumed to be a perfect electrical conductor, so that the medium is frozen onto the magnetic fieldlines. Due to the negligible gas pressure the magnetic force has nothing to oppose it so the field in the cavity is force free ($\mathbf j \times \mathbf B =0$). Hence a solution is defined by the following boundary conditions: the normal field $B_n$ at the base, the twist on each fieldline $\Phi$, and the gas pressure on the outside of the cavity $p(z)$. As the equations for the magnetic field do not involve time derivatives, the time dependent solution can be found as a series of
static solutions with differerent twists of the fieldlines at different times $t$.

Within a radius $R$ at $z=0$, a total magnetic flux $P(R,0)$ rises out of the accretion disc. The fieldlines corresponding to this $P$ turn over and return to the disc at an outer radius; axial symmetry means they define a surface of constant $P(R,z)$ as shown in figure (\ref{f4}), where $P(R,z)$ gives the flux through a circle of radius $R$ at height $z$. The disc's rotation with angular velocity $\Omega_d(R)$ produces a total twist on
the fieldlines of $\Phi(P)=\left[\Omega_d(R_i(P))-\Omega_d(R_o(P))  \right]t=\Omega(P)t$. Here $R_i(P)$ and $R_o(P)$ are the radii of the inner and outer footpoints of the field line on the flux surface $P(R,z)=P=$constant. At the boundary
of the magnetic cavity, an external gas pressure $p(z)$ is specified which must be balanced by the internal magnetic pressure. 

As $\nabla \cdot \mathbf{B}=0$ the magnetic field
may be written in terms of the flux function $P(R,z)$ as (see Mestel (1999)):
\e
\mathbf{B}=\frac{1}{2 \pi}\nabla P \times \nabla \phi+ B_\phi \hat{\phi}
\label{nodiveq}
\st
where $\phi$ is the azimuthal coordinate. With the force free condition
\e
\mathbf{j} \times \mathbf{B}=0 \Rightarrow \nabla \times \mathbf B = 4 \pi \mathbf j = \beta' \mathbf B
\label{noforceeq}
\st
this implies that taking the poloidal component of the curl of (\ref{nodiveq}) results in
\e
B_\phi=\frac{ \beta(P)}{2\pi R}; \ \ \ \ \beta'=\frac{\dd \beta}{\dd P}
\st where $\beta(P)$ is constant along each fieldline. The toroidal component leads to the differential equation for the flux function $P$:
\e
R\frac{\partial}{\partial R}\left(\frac{1}{R} \frac{\partial P}{\partial {R}} \right)+\frac{\partial^2 P}{\partial z^2}=-\beta(P)\beta'(P) \label{theeq}
\st
where the magnetic pressure must be $B^2/8\pi=p(z)$ on the surface of the magnetic cavity defined by $R=R_m(z)$.  This equation is nonlinear and is to be solved within an
unknown surface with an unknown function $\beta(P)$ (which must result in the specified values for $\Phi(P)$).

\subsection{Variational Method and its Results for the Cavity Shape} The easiest way to overcome these difficulties is to minimize the energy $W$ given by
\e
W=\int \left(\int \int \frac{B_z^2+B_R^2+B_\phi^2}{8 \pi} R \dd R \dd \phi + p(z)A \right) \dd z \st using a trial function, apply the boundary conditions and solve the resulting
variational equations as is done in paper IV. As the radial flux is constant while the magnetic cavity grows to a great height, $B_R$ is neglected. Using a trial function for the toroidal flux
such that the twist on fieldlines is the geometric mean between the case with all the twist at the top and that with the twist uniformly distributed in height, the variational equation was solved generally and the following results were obtained:
\e
\pi R_m^2=I P_m /\sqrt{2 \pi p(z)} \label{firstr} \st
\e
z^2\sqrt{8 \pi p(z)}=J^2 P_m \overline{\Phi}^2(P_m)/(16 \pi I) \label{secondr} \st where $\overline{\Phi}^2=(\Phi(P_m) / P_m)\int_{P_m}^F \Phi(P) \dd P$, $P_m(z)$ is the maximum flux $P$ at a
height $z$ and $F$ is the maximum $P$ at $z=0$. $J$ and $I$ are defined as $J=\left\langle B_\phi^2 \right\rangle /\bar{B}_\phi^2$ and $I=\left \langle B_z^2 \right \rangle / \bar{|B_z|}^2$ where $\langle \rangle$ indicates averaging over a horizontal plane, $\bar{B}_\phi=R_m^{-1}\int_0^{R_m} B_\phi \dd R$ and $\bar{|B_z|}=A^{-1} \int_0^{R_m} |B_z| 2 \pi R \dd R$; however, we approximate $J=I=1$ for our calculations (for exact values see paper IV).

The shape of the magnetic cavity can be found by solving equations (\ref{firstr}) and (\ref{secondr}) for the cavity radius $R_m(z)$. However, this only results in valid solutions when the pressure falls off more slowly than for $n=4$, where $p(z)=a(z+b)^{-n}$. (paper IV also develops a ram pressure confined model which is valid for pressure falling faster than $n=4$.) Plots of the
resulting cavity shapes show that as the jet rises it becomes more collimated; constant pressure gives conical cavities whose cross section decreases with height, deceasing pressure gives centrally bulged cavities.

The two models we will use for the relative angular velocity of the footpoints $\Omega(P)$ are
\\ i) the simple model $\Omega(P)=\Omega_0 (1-P/F)$, where $\Omega_0$ is the maximum of $\Omega$
\\ ii) the more physically realistic dipole model $\Omega(P)=\Omega_0(1-(P/F)^{3/2})$ - which corresponds to a uniformly magnetised rotating body surrounded by an accretion disc. It arises from a flux distribution $P(R,0)=FR^2/l^2$ for $R<l$ (the ``inside of the disc'') and $P(R,0)=Fl/R$ for $R>l$ (the ``outside of the disc''), combined with a footpoint angular velocity $\Omega_d=\Omega_0$ for $R<l$ and $\Omega_d=\Omega_0(R/l)^{-3/2}$ for $R>l$.

\subsection{Magnetic Field in the Cavity} 
The field within the cavity is found by writing
\e
P(R,z)=P_m(z) f(\lambda)
\label{pleq}
\st
for each height, where $\lambda=R^2/R_m^2(z)$. The resulting conditions on $f(\lambda)$ ($f(0)=f(1)=0$, $f_{\mathrm{max}}=1$) are very restrictive, so that in the tall tower approximation one can neglect variation of $f$ with height. Additionally neglecting second derivatives and squares of first derivatives of $R_m(z)$ and $P_m(z)$, equation (\ref{theeq}) becomes:
\e
\frac{4 P_m}{R_m^2} \lambda \frac{\dd^2 f}{\dd \lambda^2}=-\beta \beta'
\st

which gives $\beta \beta'$ as a product of a function of $z$ and $\lambda$. However, it is also a function of $P$, which is such a product. These facts imply that it is a power law. Now $B_\phi/B_R$ is zero on axis so $\beta(P)$ is zero when $P$ is zero. It follows that since $\beta \beta'$ is a power law, $\beta$ is as well, so we can write $\beta=C_1P^\nu $. Inserting this in equation
(\ref{theeq}) and separating variables gives:

\e
R_m=C_2 P_m^{1-\nu}
\st

\e
\lambda \frac{\dd^2 f}{\dd \lambda^2}=-C^2 \nu f^{2\nu-1} \label{flameq} \st

with $C^2=C_1^2 C_2^2/4$.

The second equation can be used to find an expression for $\nu$. Multiplying it through by $-f/\lambda$, integrating by parts, and multiplying by $P_m^2/A^2$ it can be
shown that this equation becomes
\e
\left\langle B_z^2 \right\rangle=\nu \left\langle B_\phi^2\right\rangle \st 

By expanding a slice through the jet as in Lynden-Bell (2003) with $B_R=0$, a different expression for the ratio $\left\langle B_z^2\right\rangle/\left\langle B_\phi^2
\right\rangle$ can be derived:
\e
\frac{\left\langle B_z^2\right\rangle}{\left\langle B_\phi^2 \right\rangle}=1/(2-s)
\st
where at each height
\e
s(z)=-\frac{1}{A(z) p(z)}\int_z^{\mathrm{top}} A(z')(\dd p/ \dd z') \dd z'
\st
Thus $\nu$ becomes a weak function of $z$
\e
\nu(z)=1/(2-s(z)) \st  so that the weak variation of the profile with height is determined, as a profile $f(\lambda)$ for every height can be obtained from (\ref{flameq}).

When the external pressure is constant, $f$ can be found exactly (as $\nu=1/2$). Integrating equation (\ref{flameq}) leads to the solution
\e
f(\lambda)=-e \lambda \log \lambda \label{flamc}
\st

When the pressure falls off with height, it is useful to obtain approximate analytic solutions for $f(\lambda)$ to increase calculation speed.

In the cases plotted here, the value of $\nu(z)$ is mostly $<1$. A good approximation for $\nu(z)<1$, which we will use for falling pressure throughout this paper, is $f(\lambda)=g(\lambda)/g(\lambda_1)$ where
\e
g(\lambda)=\frac{\lambda(1-\lambda^\alpha)}{\alpha(1+a_2 \lambda^\alpha)} \label{flam} \st
with $a_2$ determined by the condition that $f$ has a maximum at $\lambda=\lambda_1$
\m
a_2=(\lambda_1^\alpha(1+\alpha)-1)/(\lambda_1^\alpha((1-\alpha)-\lambda_1^\alpha))
\n
and with $\alpha(z)=2 \nu(z)-1$ and $\lambda_1(z)=(1+0.07073 \alpha(z))/e$. 
Comparing this solution to exact numerical computations shows that it is gives good results for $\nu<3$.

\section{Electromagnetic Fields}
\subsection{General Method} The jet fields must be found numerically in all but the simplest cases, as the fields are described by both implicit and differential equations. The calculations were performed using the
\emph{Mathematica} program. The errors, unless
otherwise indicated, are negligible compared with the magnitude of the results.

The different boundary conditions which define the model are
specified by the parameters $t$, $\Omega_0$, $F$ as well as,
assuming a power pressure law, $n$, $a$, $b$ where
$p(z)=a(z+b)^{-n}$. As similar jets occur in a variety of objects
with very different sizes and properties, we decided not to
attempt to choose physically realistic values for the parameters; instead we use the following default values to investigate the fields:
We choose units of time $t$ such that $2\pi$ corresponds to one turn at angular velocity $\Omega_0$ and measure magnetic flux in units of $F$. The lengthscale was then chosen so that $a=1$. With $b=0.1$ this corresponds to a radius of the disc of order $b$ for the $n=3$ case. Most plots are either for $n=3$ (which is a realistic value) or for constant pressure. Values of
$t=30$ and $b=0.1$ were used; these give a collimated tall tower where the approximations are valid. We use the simple model $\Omega(P)$ unless mentioned otherwise.

\subsection{Flux Surfaces}
\begin{figure}
				\centering
				\includegraphics[height=12cm,clip]{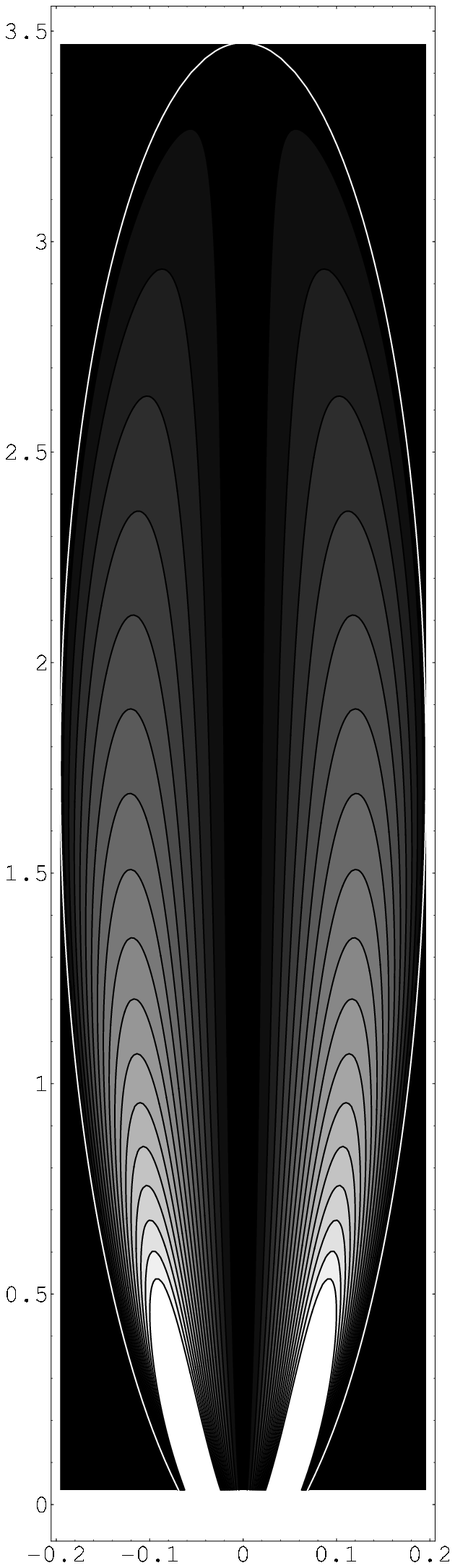}
				\includegraphics[height=12cm,clip]{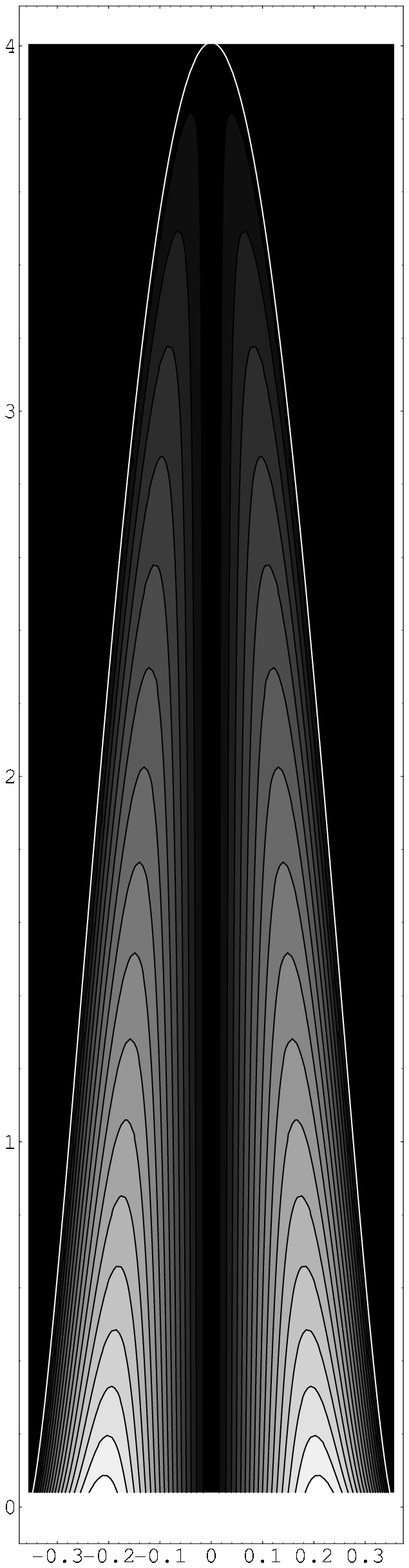}
				\caption{Cross sections of surfaces of constant flux, which also give the lines of force with the azimuthal component suppressed. On left, for $p\propto(z+0.1)^{-3}$ at $t=30$; on right, for constant $p$ at $t=90$. Both plots for are for the simple model relative rotation of footpoints. (Note: in all plots lighter shades correspond to larger values.)}
				\label{f1}
\end{figure}

As the fieldlines rise up from and return to the
disc, they define a tube-like surface of
constant flux $P$. The equation giving $P$ is
\e
P(R,z)=P_m(z) f(\lambda) =R_m^2(z)\sqrt{2 \pi p(z)} f(R^2/R_m^2(z)) /I\st
The cross sections of the surfaces of constant $P$ on which the lines of force lie are plotted in figure
(\ref{f1}) for both constant pressure and for a pressure which
falls as $n=3$. The boundary of the magnetic cavity is also
indicated in the diagrams.
\subsubsection{Results}
For the $n=3$ and $n=0$ external pressure distributions the constant flux contours / poloidal fields conform to the shape of the corresponding magnetic cavity. At the centre
of the jet, the contour lines are nearly vertical; they are closely spaced at the base.

\subsubsection{Discussion}
In the constant pressure case, it is remarkable how similar the
form of the contours and the cavity are to those in the ``dunce's
cap'' model evaluated analytically in paper IV. While
the pressure in the ``dunce's cap'' model is also constant, $\Omega(P)$ on the disc is different to the simple model used here. This
result suggests that the field structure and the shape of the jet are
determined mainly by the external pressure as the disc's rotation
profile only has a secondary influence on the jet's structure.

\subsection{Magnetic Field Components}
Substituting equation (\ref{pleq}) into the expressions for the fields given by equation (\ref{theeq}) leads to
\e
B_z(R,z)=\frac{P_m(z)}{\pi R_m^2} f'(\lambda) \st
\e
B_\phi(R,z)=\frac{C_1}{2 \pi R}(P_m(z) f(\lambda))^{\nu(z)} \st
with $P_m(z)=\pi R_m^2(z) \sqrt{2 \pi p(z)}/I$ and $\lambda=R^2/R_m^2(z)$.

\begin{figure*}
				\includegraphics[height=12cm,clip]{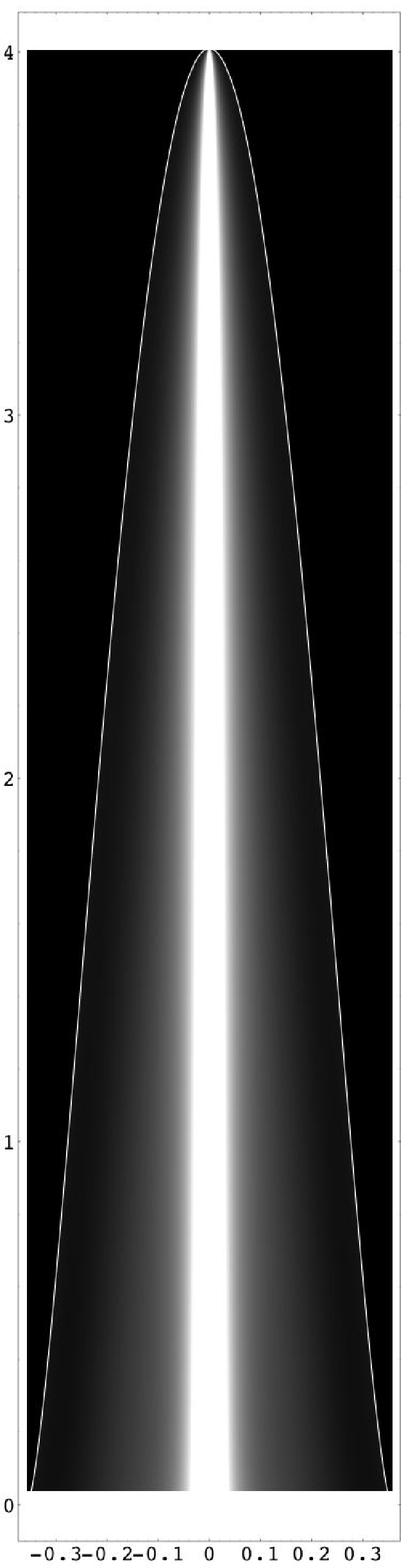}
				\includegraphics[height=12cm,clip]{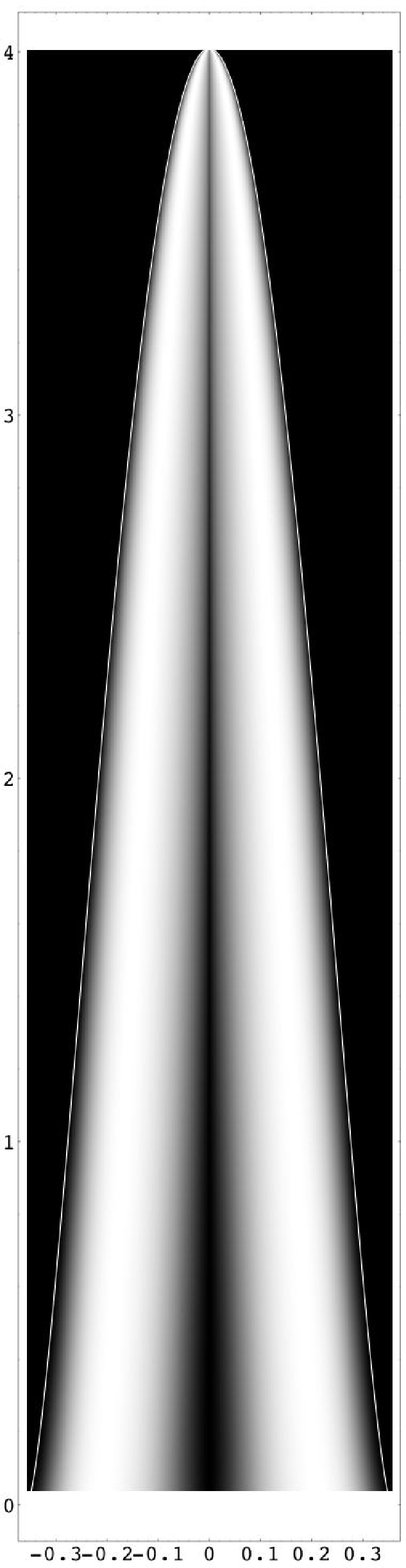}
				\caption{$B^2$ and $|B_\phi|/|B|$ for $p=$constant at $t=90$ for the simple model}
				\label{f3}
\end{figure*}

$|B_z|$ and $|B_\phi|$ were calculated for the $n=3$ pressure
law and are shown in figure (\ref{f2}). $B^2=B_\phi^2+B_z^2$ and the twist
$|B_\phi|/|B|$ were plotted for both constant pressure and the
$n=3$ pressure law in figures (\ref{f2}) and (\ref{f3})
respectively ($n=0$ plots were made at $t=50$ so that both cavities have nearly equal height). The plots are linear; to preserve detail we cut them off at large field strengths ($n=3$ cutoffs $270$, $10$, $15$ for $B^2$, $|B_\phi|$, $|B_z|$).
To investigate the dependence of the fields on $\Omega(P)$, $B^2$ was also found for the dipole model $\Omega(P)$. It is depicted in figure (\ref{f2}).

It is also useful to determine how the fields change when the
values for $t$, $\Omega_0$, $F$, $a$ and $b$ are varied. For the
$n=3$ case the parameters were individually changed from their
default values and the results noted.

\subsubsection{Results}
\begin{figure*}
				\includegraphics[height=12cm,clip]{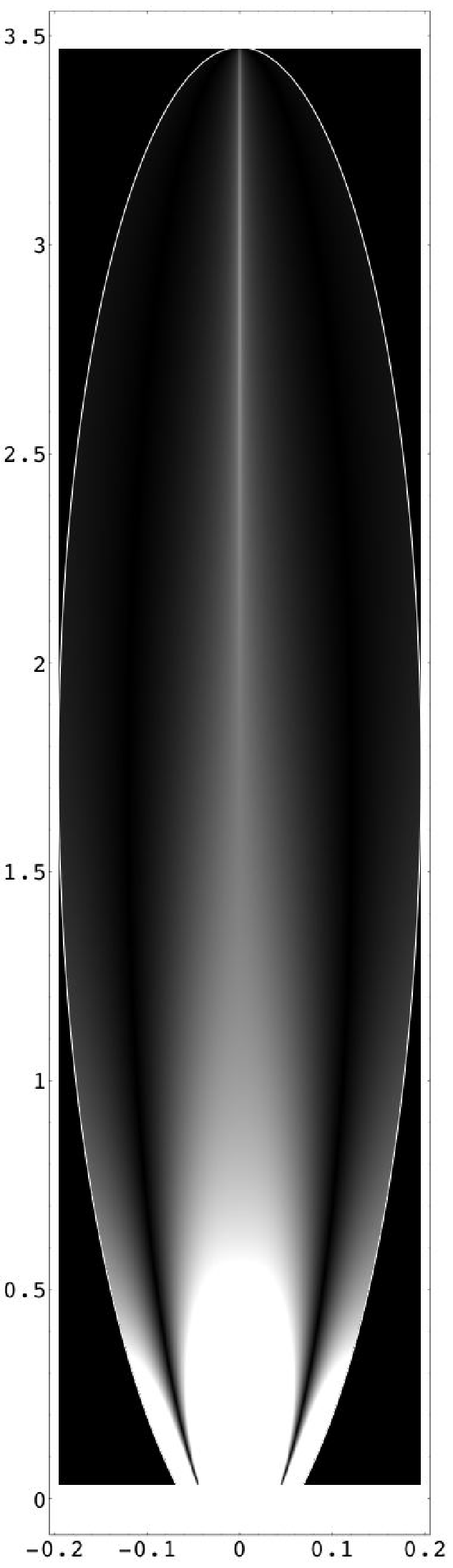}
				\includegraphics[height=12cm,clip]{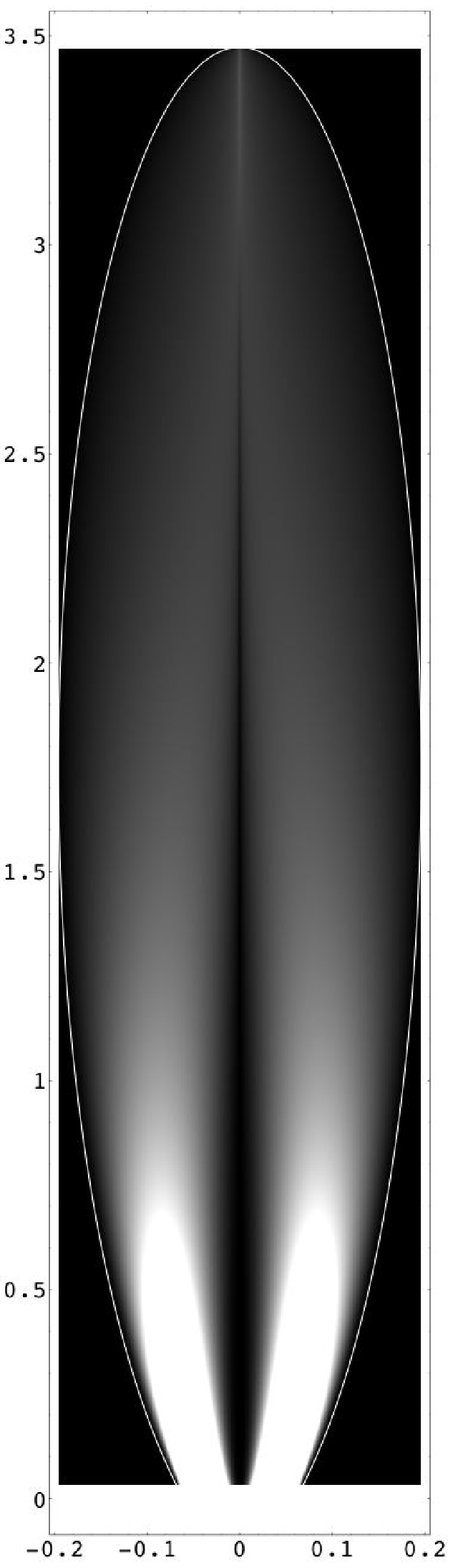}
				\includegraphics[height=12cm,clip]{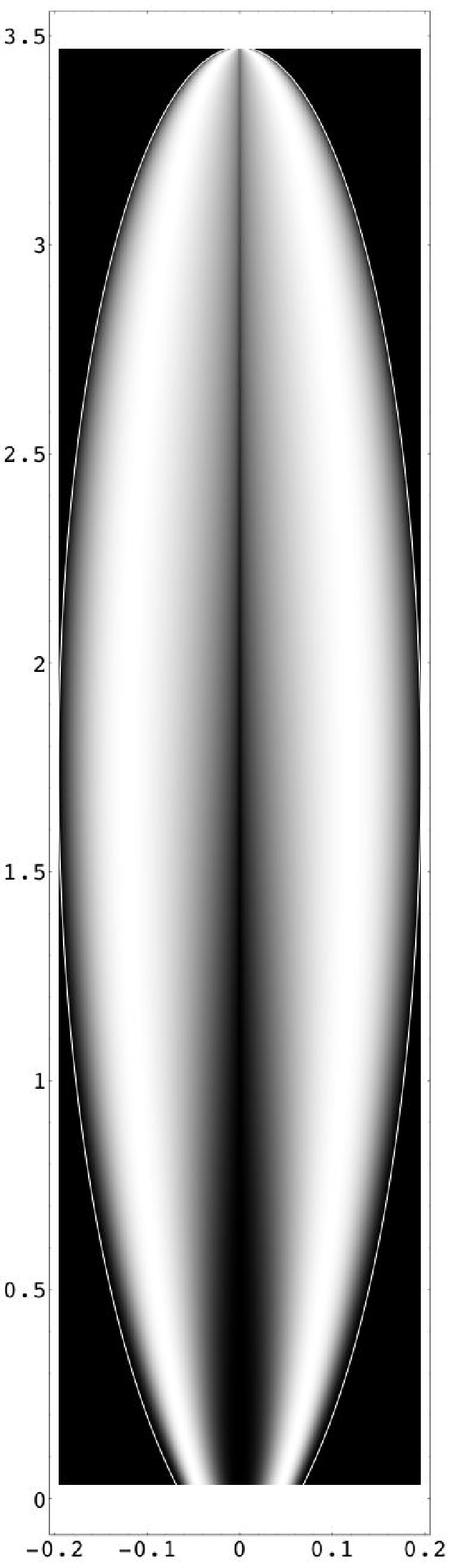}
				\includegraphics[height=12cm,clip]{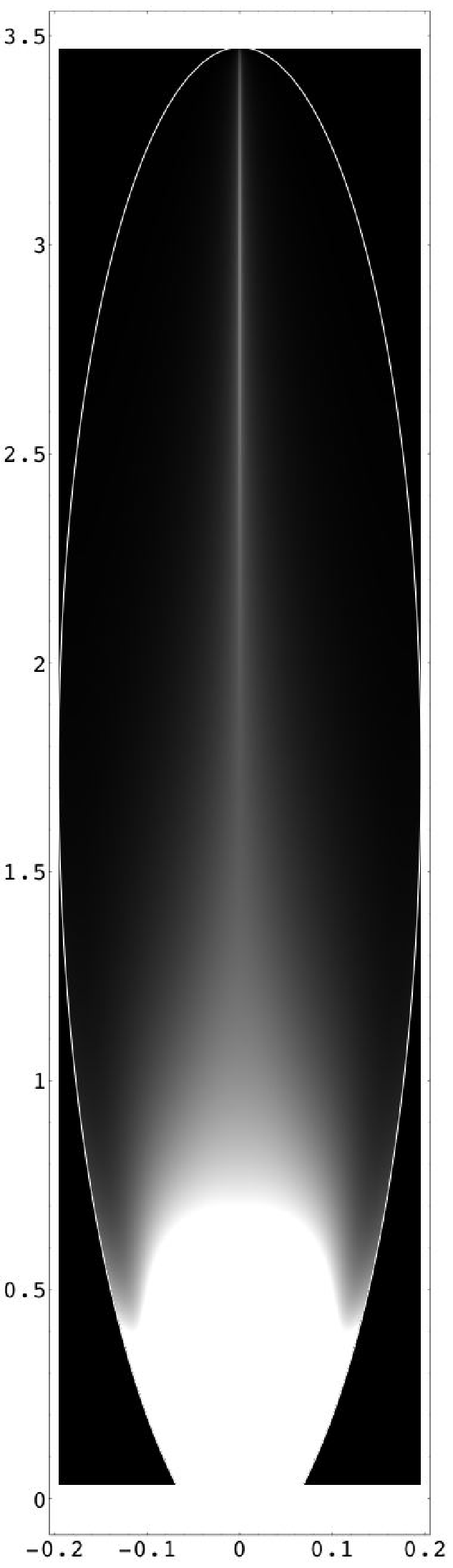}
				\includegraphics[height=11.8cm,clip]{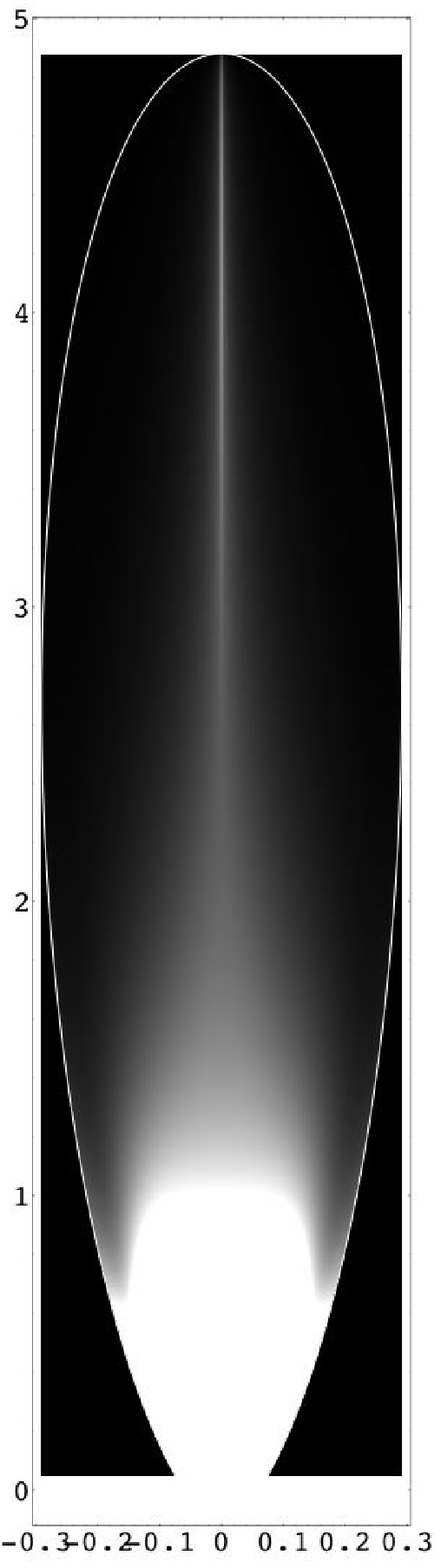}
				\caption{First four from left: plots of $|B_z|$, $|B_\phi|$, $|B_\phi|/|B|$ and $B^2$ for simple model $\Omega(P)$. For comparison, rightmost plot shows $B^2$ for dipole model disc rotation. All plots are at $t=30$ with $p\propto(z+0.1)^{-3}$}
				\label{f2}
\end{figure*}
When the pressure falls as $n=3$ it can be seen that $B_z$ is
largest at the base and core of the jet. At a radius $R_1(z)$ which corresponds to $\lambda_1$ at each height, $B_z$ goes to zero. $B_\phi$ is large at $R_1$ and at the base; there is almost no azimuthal field at the centre and at the edges of the cavity. The twist is highest at $R_1$, and is negligible on axis and near the boundary. $B^2$ is maximal at the bottom of the jet and on the upper axis.

The constant pressure jet has the highest $B^2$ field strength on axis. The field's twist is zero near the axis and the boundary.

The plot of $B^2$ for the dipole rotation model is very similar to the simple model plot; the most noticeable difference is that the dipole jet is taller.

The height and collimation of the magnetic cavities increases with time $t$, though the distribution of fields does not change much. Varying the other parameters $\Omega_0$, $F$ and $a$ changes the height of the cavity $h_{\mathrm{cav}}$ at a given time (increasing $\Omega_0$ or $F$ increases the height, increasing $a$ decreases it); however, this only
affects the field distribution substantially if the cavity is thus in a different pressure regime. So for $p(z)=a(z+b)^{-3}$, if $h_{\mathrm{cav}}$ becomes $<b$, the fields become as in the $n=0$ case, if $h_{\mathrm{cav}}\approx b$, the fields are an intermediate case, and when $h_{\mathrm{cav}}>b$ we retain the fields as in figure (\ref{f2}).

\subsubsection{Discussion}
One clearly expects strong magnetic
fields at the base because all the fieldlines rise
up from and return to the disc. When the outside pressure is $\approx$ constant, theory predicts pinching of the central axial fields by the toroidal fields, which results in a high field strength on axis, as is seen in figure (\ref{f3}). In the $n=3$ jet this pinching is only seen at the top, where $s(z)$ becomes small and $\nu(z)$ approaches $1/2$ as in the constant pressure case. The strong central fields may cause the emission of
most of the radiation in some of the observed jets.

The behaviour of $B_z$, $B_\phi$ and the twist at $R_1$ is due to the fieldlines
turning over at this radius with a large proportion of their
total twist at their tops.

The observation that the form of the fields does not depend very
strongly on $\Omega(P)$ or the other boundary conditions lets
us deduce that the plots made are quite generally valid. Ram pressure confined jet fields (see paper IV) are probably like the $n=3$ fields, as the cavity shapes are similar. The increased cavity height for the dipole model at the same $t$ is due to the larger differential rotation of the dipole model for all values of $P$.

\subsection{Individual Fieldlines}
In order to see the shapes of individual fieldlines, 3D plots were necessary. The
field lines are given by $\dd R / B_R= R \dd \phi /B_\phi = \dd z
/ B_z$, or using the expressions for the components of $\mathbf B$:
\e
\frac{\dd R}{\partial P/ \partial z}=\frac{R \dd \phi}{\beta(P)}=\frac{\dd z}{\partial P /\partial R}
\st
Thus a differential equation for the fieldline is:
\e
\frac{\dd z}{\dd \phi}=R \frac{\partial P /\partial R}{\beta(P)}=2
\lambda f'(\lambda) \frac{P_m^{1-\nu}}{C_1 f(\lambda)^\nu} \label{diff} 
\st
The rightmost expression is a function only of $z$ for a
given fieldline, i.e. a given constant value of $P$. The expression is double valued: the
fieldline rises from and then returns to the disc, so that
$\lambda(z)$ is double valued. Two different functions for
$\lambda(z)$ were obtained by solving
$f(\lambda,z)=P/P_m(z)$ for $\lambda(z)$ subject to either
$\lambda_{\mathbf{in}}(z)<\lambda_1$ or
$\lambda_{\mathbf{out}}(z)>\lambda_1$. The function $z(\phi)$ for all points on the
fieldline was found by solving equation (\ref{diff}) numerically using
$\lambda_{\mathbf{in}}(z)$ up until the point where $z$ reached
a value corresponding to the top of the fieldline, then integrating back down to the base using $\lambda_{\mathbf{out}}(z)$. The corresponding values of $R(\phi)$ were found using
$R^2=R_m^2(z(\phi)) \lambda(z(\phi))$.

\begin{figure*}
				\includegraphics[height=9cm,clip]{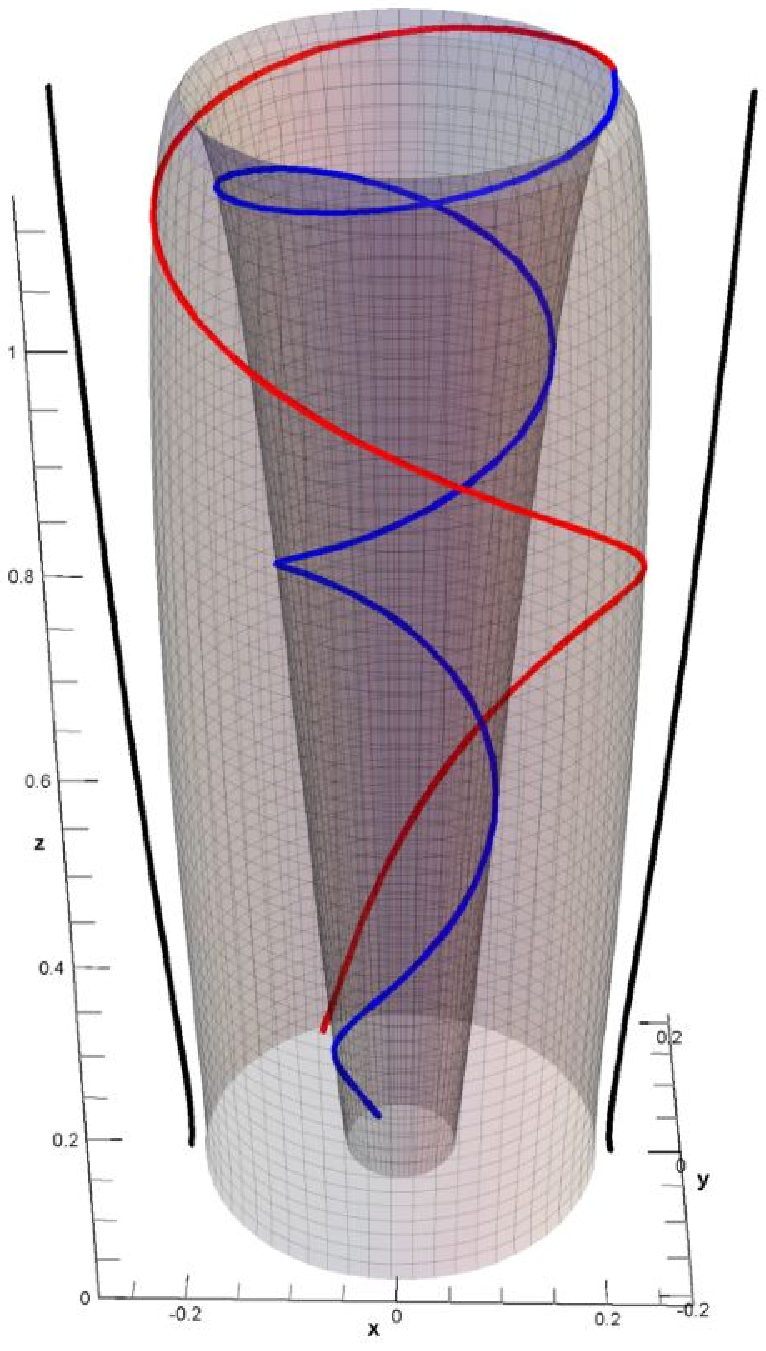}
				\includegraphics[height=9cm,clip]{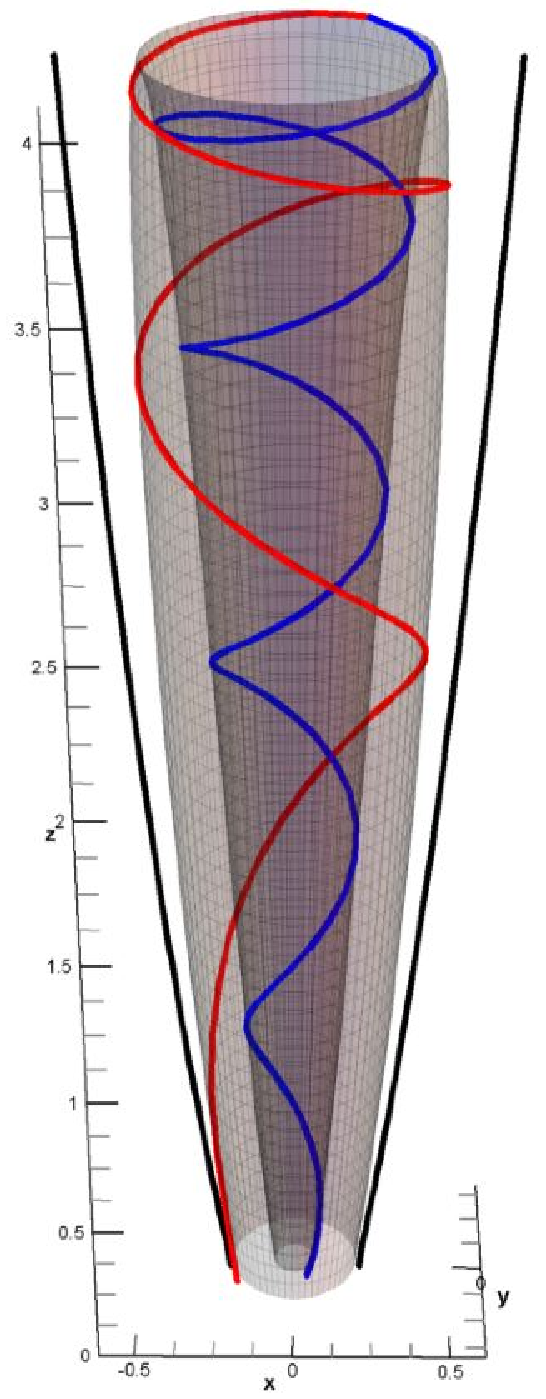}
				\includegraphics[height=9cm,clip]{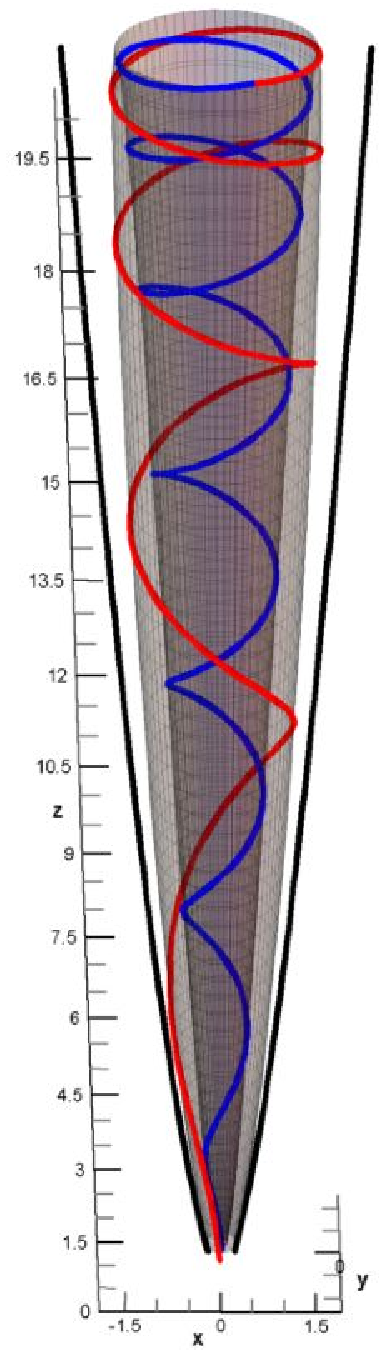}
				\caption{Individual fieldline ($P=0.3F$) at $t=30$, $t=50$, $t=90$ for simple model and $p\propto(z+0.1)^{-3}$.}
				\label{f4}
\end{figure*}

3D plots of the $P=0.3F$ fieldline (for $n=3$ and simple model $\Omega(P)$) were thus constructed at three different times. They are shown in in figure (\ref{f4}) along with the constant $P$ surface and an outline of the magnetic cavity.

In order to be able to visualize the time development of the
fieldlines, a 3D animation of the motion of the $P=0.3F$ fieldline was made for $30<t<80$. The outer foot
point of the fieldline was set to rotate with the small angular
velocity $\Omega_d=0.07$. The animation is provided on the website ``www.ast.cam.ac.uk/research/repository/bfieldline.html''.
\subsubsection{Results}
The fieldlines are twisted most at the top; the outer part of the
fieldline has less twist on it than the inner part.

The video depicts the fieldline being wound up and rising rapidly. At the bottom of the cavity the fieldline's twist decreases, so that eventually the line is nearly vertical.

\subsubsection{Discussion}
The increasing twist near the top is due to the trial function
used.
In the video it can be seen that the fieldline is dragged around by the disc at the
inside foot point and held back by the outer foot point. The
fieldline's tension thus applies a torque and transfers energy from the inner to the
outer part of the disc.

The straightening of the fieldlines near the base means that the fields become nearly antiparallel, which could lead to magnetic reconnection if the perfect conductivity breaks down.

\subsection{Poynting Flux}
\begin{figure}
\centering
				\includegraphics[height=14cm,clip]{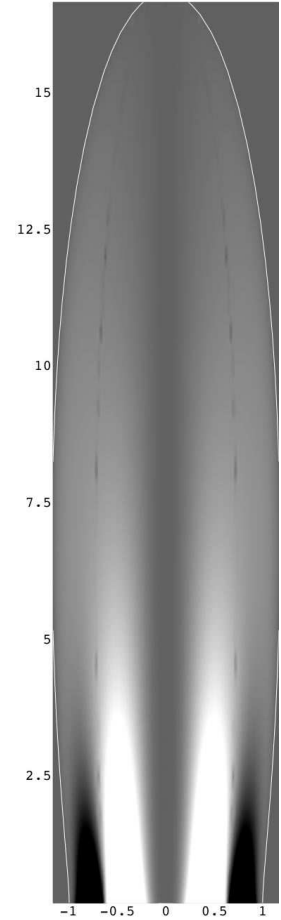}
				\caption{Plot of $R S_z$ for the dipole model at time $t=50$ with $p\propto(z+4)^{-3}$}
				\label{f5}
\end{figure}

As there is no $\mathbf{E}$ along $\mathbf{B}$ due to the assumption of perfect conductivity, and as the velocity of a fieldline $\mathbf u$ is the $c \mathbf E \times \mathbf B /B^2$ drift velocity, the electric field is given by $\mathbf{E}=-\mathbf{u} \times \mathbf{B}/c$ and $\mathbf u \cdot \mathbf B=0$. The azimuthal angle of an intersection of the fieldlines with a $z=$constant plane is shown in paper IV to be
\e
\phi=\Psi+\Omega_d(R_0) t+\Omega(P)t q(z,P)
\st
where $\Omega_d (R_0)$ is the angular velocity of the outer foot point of the field line, $\Psi$ is the initial azimuth at which the fieldline intersected the disc at the outer foot point, and $q=\int_0^z(\dd \log \lambda / \dd \log f)\dd z/\int_{\mathrm{line}}(\dd \log \lambda / \dd \log f)\dd z$ is the fraction of the total twist on the field line $P$ which occurs by height $z$ (integrating up from the outer foot point). In the same paper, the velocity of the intersection is derived in terms of the components of the fieldline's velocity $\mathbf{u}$:
\e
R \dot{\phi}=u_\phi(1+B_\phi^2/B_z^2)+u_R(B_R B_\phi/B_z^2)
\st
Applying Faraday's law to the flux through a circle about the axis in the $z=$constant plane gives
\e
2 \pi R E_\phi= -\dot{P}/c=-(u_z B_R-u_R B_z)/c\st
We solve these two equations and $\mathbf u \cdot \mathbf B = 0$ for $\mathbf{u}$; neglecting $B_R$ we obtain
\m
u_z=\frac{- B_\phi B_z R \dot{\phi}}{|B|^2};\ \ 
u_\phi=\frac{-u_z B_z}{B_\phi};\ \ 
u_R=\frac{-\dot{P}}{B_z}
\n
$\mathbf E$ follows directly from $c\mathbf{ E}=\mathbf{B} \times \mathbf{u}$, giving
\e
c E_R=-R \dot{\phi} B_z; \ \ 
c E_\phi=-\dot{P}; \ \ 
c E_z=-\dot{P} B_\phi/B_z
\label{e}
\st
It is interesting to investigate the distribution of the Poynting flux component $S_z$.
As $\mathbf S =\frac{c}{4\pi} \mathbf E \times \mathbf B$
\m
\mathbf S \propto |B|^2 \mathbf{u}
\n
so that
\e
S_z \propto - B_\phi B_z R \dot{\phi}
\st
$\dot{q}$ was approximated as $\Delta q/\Delta t$ by calculating $\Delta q$ at times separated by $\Delta t =0.03$ (using either $\lambda_{\mathrm{in}}$ or $\lambda_{\mathrm{out}}$ to overcome the double-valued-ness of $q$). Values of $t=50$ and $b=4$ were used to calculate $\mathbf S$. $\Omega(P)$ was given by the dipole model as it has a well defined $\Omega_d$. It was verified that the changes did not affect the field distributions substantially.
As calculations of $R \dot{\phi}$ were very time consuming, $B_z R \dot{\phi}$ was evaluated on a $70 \times 70$ evenly spaced grid across the width and height of the jet; cubic interpolation was then applied to obtain values at other points. The vertical Poynting flux was thus plotted in figure (\ref{f5}).
\subsubsection{Results}
The $z$ component of the Poynting flux is large and positive just above the inside of the accretion disc. Above the outer part of the disc there is a large downward flow.
\subsubsection{Discussion}
Energy flows up from the inside of the disc because the disc drags the fieldlines around and thus does work on the field. The flow spreads to the top and sides of the jet, where it enables the expansion against the external pressure. Some of the energy returns to the outer part of the disc (as the fields pull on the disc there); the fraction this constitutes was calculated as $\int_{R_1}^{R_m(0)} R S_z \dd R/\int_0^{R_1} R S_z \dd R=0.47$. It thus appears that jets are quite an efficient mechanism for energy transfer from the inside to the outside of the accretion disc. The energy flow into the jet can also be calculated by considering torques on the disc. The couple $G \dd P=-\frac{B_z B_\phi}{4 \pi} 2 \pi R \dd P$ carried by the tube of flux between $P$ and $P+ \dd P$ is constant along the tube. This couple inevitably does work on the rotating outer feet of the field lines, giving a power flow $G(P) \dd P \ \Omega(R_o(P))$ down from the jet into the outside of the disc. This differs from the results of many simulations (Ustyugova et al. (1999), Gammie et al. (2003), McKinney (2005), McKinney and Narayan (2007a/b) and Narayan et al. (2007) on winds) because they have no magnetic flux returning to the disc and so do not have the downward Poynting flux which results from this. Romanova et al. (1998) do not have a force-free magnetosphere. The disc does work on the field lines at the inner feet, giving a power input of $G(P) \dd P \ \Omega(R_i(P))$. The difference between the power input and output goes into expanding the magnetic cavity against the pressure and increasing the magnetic energy. Integrating over all flux tubes, the net power put into the expanding jet is $\int_0^F G(P) \Omega(P) \dd P$.

As $-\dot{P}/c= E_\phi$ is small compared to the other $\mathbf E$ components but $\mathbf E \bot \mathbf B$, $\mathbf E$ points close to normally out of the $P$ surface at each point.

\section{Conclusions}
1. The $\mathbf B$ field is strongest at the base of the magnetic cavity and on axis. The field structure is only weakly affected by the detailed profile of the shear in the accretion disc's rotation.\\
2. The fieldlines are twisted more at their tops and at smaller radii. At the cavity base the fieldlines become straight and vertical.\\
3. In the case investigated the energy rises from the inside of the disc; somewhat less than half returns to the outer part of the disc, the remainder flows up the jet.

The published numerical calculations of others referred to above do not extend to great heights due to the difficulties with numerically induced resistivity. Although in our work only the ``dunce's cap'' solution is truly exact, the solutions in this paper are still basically analytic and do not suffer this inconvenience. In the few cases such as Nakamura, Li and Li (2006) where similar problems are solved, the results agree with the start of our solutions.

\section{Acknowledgements}
B.D.S. would like to acknowledge a grant from the Nuffield Foundation which supported this work.


\begin{thebibliography}{90}
\bibitem{}
Baade W., 1956, ApJ, 439, L43
\bibitem{}
Barnes C.W., Sturrock P.A., 1972, ApJ, 174, 659
\bibitem{}
Begelman M.C., Blandford R.D., Rees M.J., 1984, Rev. Mod. Phys., 56, 255
\bibitem{}
Carilli C.L., Barthel P.D., 1996, A\&ARv, 7, 1
\bibitem{}
Gammie C.F., McKinney J.C., G\'abor T., 2003, ApJ, 589, 444
\bibitem{}
Gizani N.A.B., Leahy J.P., 2003, MNRAS, 342, 399
\bibitem{}
Hargrave P.J., Ryle M., 1974, MNRAS, 166, 305
\bibitem{}
Kato Y., Mineshinge S., Shibata K., 2004, ApJ, 605, 307
\bibitem{}
Li H., Lapenta G., Finn J.M., Li S., Colgate S.A., 2006, ApJ, 643, 92
\bibitem{}
Lovelace R.V.E., 1976, Nat, 262, 649
\bibitem{}
Lovelace R.V.E., Li H., Koldoba A.V., Ustyugova G.V., Romanova M.M., 2002, ApJ, 572, 445
\bibitem{}
Lynden-Bell D., 1996, MNRAS, 279, 389 (paper II)
\bibitem{}
Lynden-Bell D., 2003, MNRAS, 341, 1360 (paper III)
\bibitem{}
Lynden-Bell D., 2006, MNRAS, 369, 1167 (paper IV)
\bibitem{}
McKinney J.C., 2005, ApJ, 630, L5
\bibitem{}
McKinney J.C., Narayan R., 2007a, MNRAS, 375, 513
\bibitem{}
McKinney J.C., Narayan R., 2007b, MNRAS, 375, 531
\bibitem{}
Mestel L., 1999, Stellar Magnetism, pp. 53-70, OUP
\bibitem{}
Nakamura M., Meier D.L., 2004, ApJ, 617, 123
\bibitem{}
Nakamura M., Li H., Li S., 2006, ApJ, 652, 1059
\bibitem{}
Nakamura M., Li H., Li S., 2006, preprint astro-ph/0609007
\bibitem{}
Narayan R., McKinney J.C., Farmer A.J., 2007, MNRAS, 375, 548
\bibitem{}
Romanova M.M., Ustyugova G.V., Koldoba A.V., Chechetkin V.M., Lovelace R.V.E., 1998, ApJ, 500, 703
\bibitem{}
Sturrock P.A., Barnes C.W., 1972, ApJ, 176, 31
\bibitem{}
Ustyugova G.V., Koldoba A.V., Romanova M.M., Chechetkin V.M., Lovelace R.V.E., 1999, ApJ, 516, 221
\bibitem{}
Uzdensky D.A., MacFadyen A.I., 2006, ApJ, 647, 1192
\bibitem{}
Vlemmings W.H.T., Diamond P.J., Imai H., 2006. Nat, 440, 58
\end{thebibliography}
\end{document}